\documentclass[aps,prb,superscriptaddress,epsfigm,preprint,showpacs]{revtex4}
\usepackage{graphicx}
\usepackage{amsmath}

\begin{document}

\title{Phonon-induced electron relaxation in weakly-confined single and coupled quantum dots}

\author{J.I. Climente}
\affiliation{CNR-INFM S3, Via Campi 213/A, 41100 Modena, Italy}
\email{climente@unimo.it}
\homepage{http://www.nanoscience.unimo.it/}
\author{A. Bertoni}
\affiliation{CNR-INFM S3, Via Campi 213/A, 41100 Modena, Italy}
\author{G. Goldoni}
\affiliation{CNR-INFM S3, Via Campi 213/A, 41100 Modena, Italy}
\affiliation{Dipartamento di Fisica, Universit\`a degli Studi di Modena e Reggio Emilia, 
Via Campi 213/A, 41100 Modena, Italy}
\author{E. Molinari}
\affiliation{CNR-INFM S3, Via Campi 213/A, 41100 Modena, Italy}
\affiliation{Dipartamento di Fisica, Universit\`a degli Studi di Modena e Reggio Emilia, 
Via Campi 213/A, 41100 Modena, Italy}
\date{\today}

\begin{abstract}
We investigate charge relaxation rates due to acoustic phonons
in weakly-confined quantum dot systems, including both deformation 
potential and piezoelectric field interactions. 
Single-electron excited states lifetimes are
calculated for single and coupled quantum dot structures, both in
homonuclear and heteronuclear devices. Piezoelectric field
scattering is shown to be the dominant relaxation mechanism in many
experimentally relevant situations. On the other hand, we show that
appropriate structure design allows to minimize separately
deformation potential and piezolectric field interactions, and may
bring electron lifetimes in the range of microseconds.
\end{abstract}


\maketitle
\section{Introduction}

Single-electron excited states lifetimes in GaAs-based quantum dots
(QDs) are limited to the order of
nanoseconds.\cite{FujisawaNAT,FujisawaJPCM} This limit is set mainly
by acoustic-phonon scattering, which has been shown to be the
dominant scattering mechanism between discrete energy states with
few-meV gap in both single\cite{FujisawaNAT,FujisawaJPCM} and
vertically coupled\cite{FujisawaSCI, OrtnerPRB} quantum dots (SQDs
and CQDs). Prediction of, and control over the electron relaxation
rates is desirable for many possible QD applications, ranging from
decoherence-free implementation of quantum gates\cite{ZanardiPRL} to 
efficient QD lasers.\cite{SugawaraAPL,WingreenIEEE} It is also desirable 
in order to achieve high-resolution optical spectroscopy, since this requires
that (non-radiative) phonon scattering rates are smaller than
(radiative) photon emission and absorption rates. Therefore, the
knowledge of the physics of electron coupling to acoustic phonons in
these nanostructures is of great interest.

In a pioneering work, Bockelmann investigated theoretically the
influence of lateral (spatial and magnetic) confinement on the
single-electron relaxation rates in parabolic
QDs.\cite{BockelmannPRB} To this end, he considered the deformation
potential (DP) coupling between electrons and longitudinal acoustic
phonons, while neglecting piezoelectric (PZ) coupling on the grounds
of its weaker contribution in 2D GaAs/Ga$_x$Al$_{1-x}$As structures,
and the fact that a similar qualitative behavior could be expected
in QDs. Using the same assumption, it has been recently proposed
that orders-of-magnitude suppression of phonon-induced scattering
rates can be achieved \cite{ZanardiPRL,BertoniAPL,BertoniPE} in
properly designed CQD structures, and the effect of
electron-electron interaction in multi-electron QD structures has
been investigated.\cite{BertoniPRL}

On the other hand, recent theoretical and experimental works suggest
that electron-acoustic phonon scattering due to PZ field interaction
is indeed relevant for momentum and spin-relaxation processes in
QDs,\cite{FujisawaJPCM,ChengPRB} and may even provide the leading
contribution to charge decoherence in laterally coupled
QDs.\cite{WuPRB,StavrouPRB} Therefore, the use of a theoretical
model which considers simultaneously both DP and PZ coupling of
electrons to acoustic phonons in QDs appears necessary to evaluate
quantitatively electron relaxation rates and assess to which extent
previous predictions are affected by the PZ field.

In this paper we study the phonon-induced single-electron relaxation
rates in realistic models of weakly-confined single and vertically 
coupled QDs, taking into account both DP and PZ mechanisms. The contribution of
each relaxation channel is singled out, and the regimes where each
coupling mechanism prevails are established. The PZ coupling is
shown to be the prevailing one in several experimental situations.
Furthermore, we show that QD devices with maximum lifetimes, in the
range of microseconds, can be realized not only in coupled QDs, as
previously predicted,\cite{ZanardiPRL,BertoniAPL} but also in single
QDs.

\section{Theoretical considerations}

The theoretical model we use is essentially the same of our previous
works,\cite{BertoniAPL,BertoniPE} but now including the scattering
rate arising from the PZ field due to longitudinal-acoustic (LA) and
transverse-acoustical (TA) phonons. We study GaAs/AlGaAs dots with
disk shape and confinement energy in the range of few meV.\cite{FujisawaNAT,FujisawaJPCM,weakdots} 
The confinement potential has been modeled as a quantum well in the
growth direction $z$, formed by the heterostructure band-offset,
while in the $xy$ plane a 2D parabolic confinement is assumed, which
gives rise to the Fock-Darwin level structure.\cite{ReimannRMP} The
three-dimensional single-particle electronic states are computed
within the envelope function approximation.  The electron states are
labelled by the three quantum numbers $(n,m,g)$, where
$n=0,1,2\ldots$ denotes the $n$-th state with azimuthal angular
momentum $m=0,\,\pm 1,\,\pm 2\ldots$ and parity $g$. Here $g=0$
($1$) stands for
even (odd) parity with respect to the reflection about the $z=0$ plane. 
We consider bulk phonons with linear dispersion $\omega_{\sigma q}=
c_{\sigma} \lvert \mathbf{q} \rvert$, where $c_{\sigma}$ is the
sound velocity of LA ($\sigma=\mbox{LA}$) or TA ($\sigma=\mbox{TA}$)
phonon modes in the QD material, and $\mathbf{q}$ is the momentum.\cite{approx}
The electron-phonon interaction Hamiltonian reads ${\cal
H}_{e-p}=\sum_{\nu \mathbf{q}}
\,M_{\nu}(\mathbf{q})\,\left(b_{\mathbf{q}}\,e^{i \mathbf{q r}} +
b_{\mathbf{q}}^\dag\, e^{-i\,\mathbf{q r}}\right)$, where
$b_{\mathbf{q}}$ and $b_{\mathbf{q}}^\dag$ are the phonon
annihilation and creation operators respectively, and
$M_{\nu}(\mathbf{q})$ is the scattering matrix element corresponding
to the electron scattering mechanism $\nu$ (see below).  For
simplicity we assume zero temperature, so that phonon absorption and
multi-phonon processes are neglegible. The relaxation rate between
the initial (occupied) state $\lvert \Psi_i\rangle$ and the final
(unoccupied) state $\lvert \Psi_f\rangle$ is determined by Fermi
golden rule:

\begin{equation}
\tau^{-1}_{if}=\frac{2\pi}{\hbar}\,\sum_{\nu \mathbf{q}} \lvert M_{\nu}(\mathbf{q})\rvert^2\,
\lvert\langle \Psi_f|\,e^{-i \mathbf{q r}}\,|\Psi_i \rangle\rvert^2\,
\delta(|E_f-E_i| - E_q),
\label{eq1}
\end{equation}

\noindent where $E_f$ and $E_i$ stand for the final and initial electron states energy and
$E_q=\hbar \omega_{\sigma q}$ represents the phonon energy. It is clear from the above equation that the
relaxation is mediated by phonons whose energy matches that of the transition between the initial
and final electron states.  The electron-phonon scattering is composed of the following
contributions:\cite{Gantmakher_book,ZookPR}

\noindent (i) The electron-LA phonon scattering due to the deformation potential
($\nu=$LA-DP):

\begin{equation}
\lvert M_{\mbox{\tiny LA-DP}}(\mathbf{q}) \rvert^2=\frac{\hbar\,D^2}{2\,d\,c_{\mbox{\tiny LA}}\,\Omega}\,|\mathbf{q}|,
\label{eq2}
\end{equation}

\noindent (ii) the electron-LA phonon scattering due to the PZ field
($\nu=$LA-PZ):

\begin{equation}
\lvert M_{\mbox{\tiny LA-PZ}}(\mathbf{q}) \rvert^2=\frac{32 \pi^2\,\hbar\,e^2\,h_{14}^2}{\epsilon^2\,d\,c_{\mbox{\tiny LA}}\,\Omega}\,
\frac{(3\,q_x\,q_y\,q_z)^2}{|\mathbf{q}|^7},
\label{eq3}
\end{equation}

\noindent (iii) the electron-TA phonon scattering due to the PZ
field ($\nu=$TA-PZ):

\begin{equation}
\label{eq4}
\lvert M_{\mbox{\tiny TA-PZ}}(\mathbf{q}) \rvert^2=\frac{32 \pi^2\,\hbar\,e^2\,h_{14}^2}{\epsilon^2\,d\,c_{\mbox{\tiny TA}}\,\Omega}
\left| \frac{q_x^2\,q_y^2 + q_y^2\,q_z^2 + q_z^2\,q_x^2}{|\mathbf{q}|^5} - \frac{(3\,q_x\,q_y\,q_z)^2}{|\mathbf{q}|^7} \right|.
\end{equation}


In the above expressions $D$, $d$ and $\Omega$ stand for the crystal
acoustic deformation potential constant, density and volume,
respectively. $e$ is the electron charge, $h_{14}$ the PZ constant
and $\epsilon$ the static dielectric constant. Note that the sum in
Eq.~(\ref{eq1}) runs twice over $\nu=$TA-PZ, to account for the two
transverse phonon modes (both identical under
Ref.~\onlinecite{ZookPR} approximation for zinc-blende crystals).
For the analysis of the results presented in the following sections,
it is useful to realise that differences between the various
scattering mechanisms in Eq.~(\ref{eq1}) arise from i) the diverse
matrix elements $M_{\nu}(\mathbf{q})$; ii) the different value of
$\mathbf{q}$ associated with a given transition energy for TA and LA
phonons which enters $\lvert\langle \Psi_f|\,e^{-i \mathbf{q r}}\,
|\Psi_i \rangle\rvert^2$. The former factor introduces some
qualitative differences in the behavior, whereas the latter mostly
shifts the transition energy values at which LA- and TA-phonon
scattering with the same phonon momentum occur.

In this work we consider mostly relaxation rates from the first
excited to the ground electron state in SQDs and CQDs. This is often
the most relevant transition, since many QD applications rely on the
formation of two-level systems, and it can be monitored e.g. by
means of pump-and-probe techniques.\cite{FujisawaNAT,FujisawaJPCM}
Assuming Fock-Darwin level structure, for SQDs this corresponds to a
transition from a $p$ ($m=1$) to a $s$ ($m=0$) level. For CQDs this
corresponds to the isospin transition from an antisymmetric ($g=1$)
to a symmetric ($g=0$) state. The behavior of transition rates from
higher excited states is qualitatively similar, except for the
presence of a larger number of decay channels which smears the
features of direct scattering between two selected
states.\cite{BockelmannPRB} We use GaAs/Al$_{0.3}$Ga$_{0.7}$As
material parameters:\cite{Tin_book} electron effective mass
$m^*=0.067$, band-offset $V_c=243$ meV, $d=5310$ kg/m$^3$, $D=8.6$
eV, $\epsilon=12.9$, and $h_{14}=1.41\cdot 10^9$ V/m. For the sound
speed $c_{\sigma}$, we take into account that in cylindrical QDs
most of the scattering arises from phonon propagation close to the
growth direction.\cite{BertoniPE} We then assume that the QDs are
grown along the $[1\, 0\, 0]$ direction and use the corresponding
values $c_{\mbox{\tiny LA}}=4.72 \cdot 10^3$ m/s and $c_{\mbox{\tiny
TA}}=3.34 \cdot 10^3$ m/s.\cite{Landolt_book}

\section{Relaxation in Single Quantum Dots}

In this section we study the $(n,m,g)=(0,1,0) \rightarrow (0,0,0)$ transition in
SQDs as a function of the spatial and magnetic confinement.

\begin{figure}[p]
\includegraphics[width=8.5cm,clip]{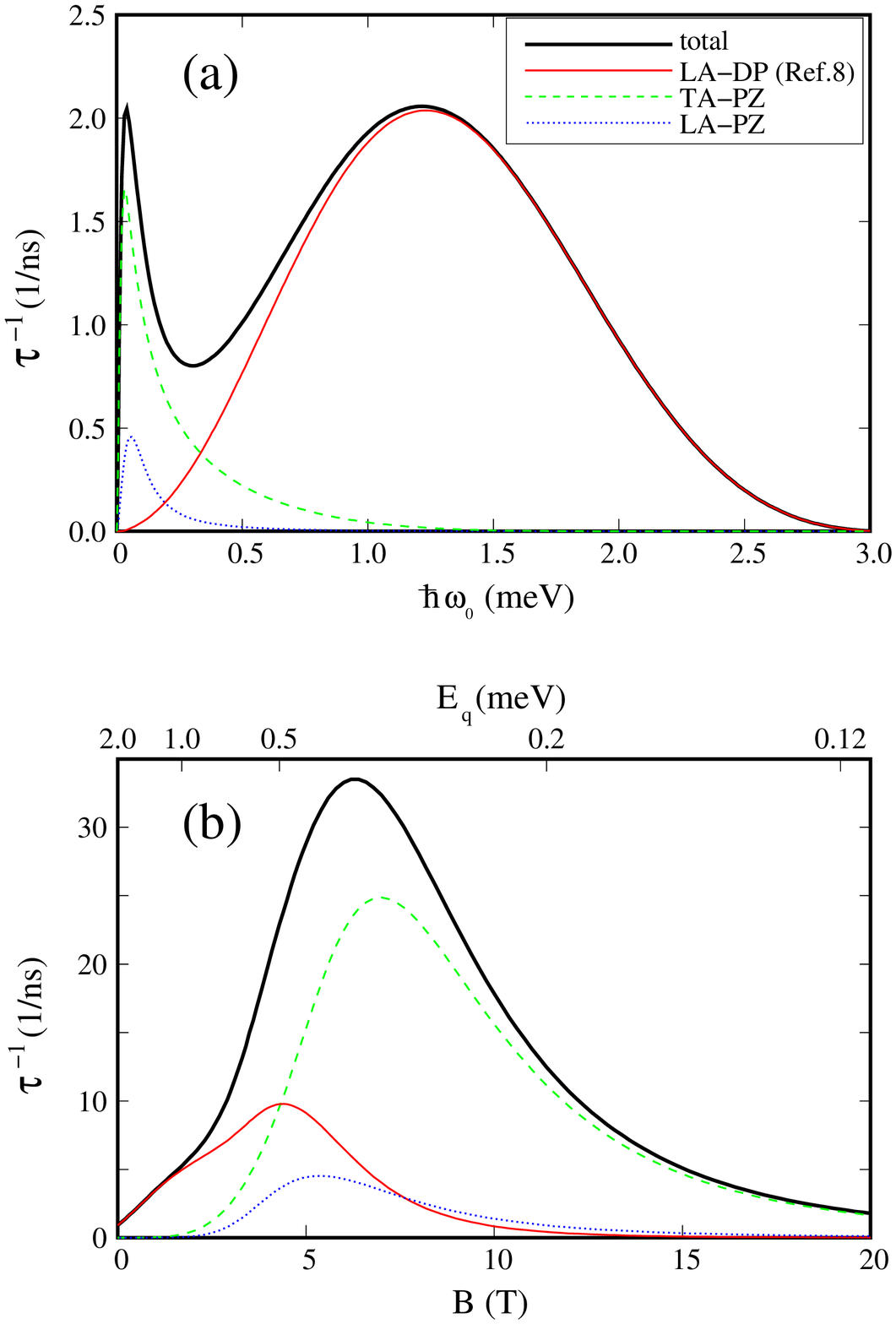}
\caption{(Color online) Acoustic phonon scattering rate as a function of (a)
lateral confinement, and (b) magnetic field, for the electron
transition $(n,m,g)=(0,1,0) \rightarrow (0,0,0)$ in a
GaAs/Al$_{0.3}$Ga$_{0.7}$As QD with $L_z=10$ nm. In panel (b) $\hbar
\omega_0=2$ meV, and the upper scale shows the emitted phonon energy
$E_q$.}\label{Fig1}
\end{figure}

Figure \ref{Fig1}(a) illustrates the scattering rate of a SQD with well width
$L_z=10$ nm as a function of the harmonic confinement energy, $\hbar \omega_0$.
 For most lateral confinements, DP coupling (the only source of relaxation considered
in Ref.~\onlinecite{BockelmannPRB}) gives the largest contribution.
However, for weak confinements ($\hbar \omega_0 < 0.4$ meV) PZ
coupling prevails, so that the total scattering rate shows two
maxima instead of one. This is because the different wave vector
dependence in the DP and PZ matrix elements ($\sqrt{q}$ for DP
versus $1/\sqrt{q}$ for PZ interaction) yields maxima at different
confinement energies (i.e., different energies of the emitted
phonons). Moreover, we observe that the PZ contribution coming from
TA-phonons is larger than that of LA-phonons.  This holds for almost
all calculations throughout this paper. Although the shape of PZ
scattering rate is different from the DP one, the limiting behavior
is similar: it tends to zero at very weak confinement potentials
because of the decreasing phonon density of states; it also tends to
zero in the strong confinement limit, due to the orthogonality of
the electron states which makes the factor $\langle \Psi_f|\,e^{-i
\mathbf{q r}}\,| \Psi_i \rangle$ vanish rapidly when the phonon
wavelength $\lambda_q$ is shorter than the lateral (i.e., the
largest) confinement length.\cite{BockelmannPRB,BertoniPE}

The lateral confinement and transition energy in a QD can be also
modulated by a magnetic field ($B$) applied along the growth
direction. In Figure \ref{Fig1}(b) we show the $B$-dependence of the
scattering rate for a SQD with $L_z=10$ nm and $\hbar \omega_0=2$
meV. We note that the PZ rate is neglegible at zero magnetic field,
but it rapidly increases with $B$ and soon exceeds the DP rate. This
is because the energy of the (0,1,0) and (0,0,0) states converge
with increasing $B$,\cite{ReimannRMP} so that the energy
of the emitted phonon becomes ever smaller (see upper scale in the
figure) and the PZ matrix elements are maximized. As a consequence,
the maximum of the total scattering rate, including the PZ
interaction, is over three times higher than the one
calculated including only the DP mechanism; this difference is even
larger for dots with weaker lateral confinement (not shown).
Therefore, the inclusion of PZ interactions to properly describe the
scattering rate in SQDs in the presence of magnetic fields is
critical. This result also justifies recent findings reported by
other authors, which claim that spin relaxation rate is mostly
determined by TA-PZ scattering:\cite{ChengPRB} spin-orbit mixing is
only significant when $B$ drives the electron states involved close
in energy\cite{FlorescuPE}, and then TA-PZ is by far the most
important source of phonon scattering. One may note in
Figure \ref{Fig1}(b) that the maxima of TA-PZ and LA-PZ scattering
rates take place at different values of $B$, in spite of the fact
that the matrix elements of both mechanisms depend on the phonon
wave vector as $1/\sqrt{q}$. This is because of the different sound
velocities of TA and LA phonon modes, which associate the same
phonon energy with different wave vectors,
$|\mathbf{q}|= E_q /(\hbar c_{\sigma})$.\\

\section{Controlling electron relaxation}
\subsection{Single Quantum Dots}

\begin{figure}[p]
\includegraphics[width=8.5cm,clip]{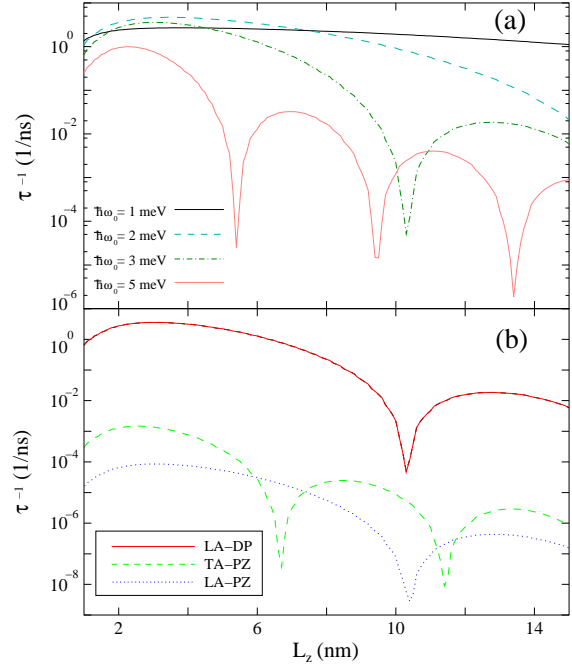}
\caption{(Color online) (a) Acoustic phonon total scattering rate as a function of
the quantum well width $L_z$ for the electron transition
$(n,m,g)=(0,1,0) \rightarrow (0,0,0)$ in a SQD, at selected values
of lateral confinement $\hbar\omega_0$. (b) Individual contributions
from each scattering mechanism in a QD with $\hbar \omega_0=3$
meV.}\label{Fig2}
\end{figure}

Electron-acoustic phonon interaction in SQDs is to a large extent
determined by the interplay of lateral confinement length, quantum
well width and phonon wavelength.\cite{BockelmannPRB} However, in
weakly-confined QDs vertical confinement is usually much stronger
than lateral confinement. As a result, the energy of the transition
$(0,1,0) \rightarrow (0,0,0)$ (and thus the wavelength of the
emitted phonon) is exclusively determined by the lateral
confinement. Therefore, if we fix the lateral confinement and change
only the well width we could expect that the scattering rate
exhibits periodic oscillations, with maxima when $L_z \simeq (j+1/2)
\lambda_q$ (i.e., the electron wave function along the quantum well
direction is in-phase with the phonon wave) and minima when $L_z
\simeq j \lambda_q$ (i.e., the electron wave function is in
anti-phase with the phonon wave). We explore this possibility in
Figure \ref{Fig2}(a), where the effect of vertical confinement on the
total scattering rates of SQDs with different $\hbar \omega_0$ is
illustrated. For $\hbar \omega_0=1$ and $\hbar \omega_0=2$ meV no
scattering minima are observed, indicating that $\lambda_q$ is still
too large for the range of $L_z$ shown (which are indeed typical
well widths of realistic devices). However, for $\hbar \omega_0=3$
meV we observe a striking dip at about $L_z=10.5$ nm, where the
scattering rate is suppressed by orders of magnitude due to the
anti-phase relation between the electron wave function and the
phonon plane wave. For $\hbar \omega_0=5$ meV order-of-magnitude 
damping of the scattering rate at several well widths values is observed.
A similar geometry-induced suppression of the scattering
rate was previously proposed for CQDs\cite{ZanardiPRL}, but here we
show that it is also feasible for SQDs with realistic dimensions if
the confinement energy is in the proper range. This result may be of
great significance for quantum coherence preservation in SQD
devices.

Figure \ref{Fig2}(b) depicts separately the LA-DP, TA-PZ and LA-PZ
scattering rates in the SQD with $\hbar \omega_0=3$ meV. One can see
that, in the absence of external fields, LA-DP scattering rate
constitutes the dominant relaxation mechanism by at least one order
of magnitude. This accounts for the fact that only one damping dip
is observed in the total scattering rate, even though TA-PZ
scattering shows two dips due to the smaller speed of TA phonons.

\begin{figure}[p]
\includegraphics[width=8.5cm,clip]{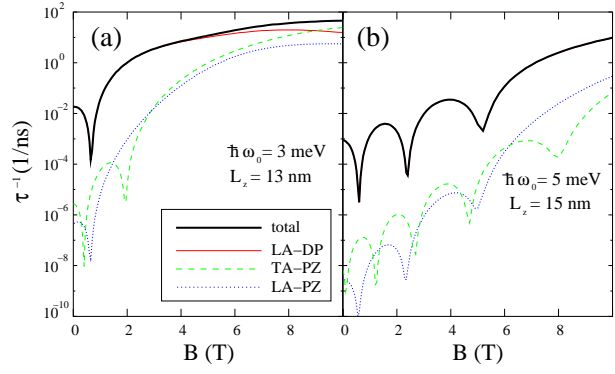}
\caption{(Color online) Acoustic phonon scattering rate as a function of the
magnetic field for the electron transition $(n,m,g)=(0,1,0)
\rightarrow (0,0,0)$ in selected SQDs.}\label{Fig3}
\end{figure}

From a practical point of view, it is clearly difficult to fabricate
QDs with the exact aspect ratio (i.e., lateral vs vertical
confinement) which meets the scattering rate minima plotted in
Figure \ref{Fig2}. Therefore, we next show that the damping of
phonon-induced relaxation rate in SQDs can be also achieved by means
of external magnetic fields. Let us consider a SQD of lateral
confinement $\hbar \omega_0$ and well width $L_z > L_z^{crit}$,
where $L_z^{crit}$ is a well width value leading to a scattering
rate minimum.
A magnetic field applied along the growth direction decreases the
energy spacing between the $(0,1,0)$ and $(0,0,0)$ states and, as a
consequence, increases the wavelength of the emitted phonon. It is
then possible to tune $B$ in order to obtain the phonon wavelength
which is in anti-phase with the electron wave function in the
quantum well of width $L_z$. This is illustrated in
Figure \ref{Fig3}(a) for a SQD with $\hbar \omega_0=3$ meV and $L_z=13$
nm, and in Figure \ref{Fig3}(b) for a SQD with $\hbar \omega_0=5$ meV
and $L_z=15$ nm. In both cases scattering rate dips similar to those
observed in Figure \ref{Fig2} are forced by the magnetic field.
Moreover, the total scattering rate suppressions are again well
described in terms of DP coupling,
 since PZ coupling becomes significant at larger values of $B$ only.
This indicates that, as a general rule, for usual QD heights, phonon energies leading to scattering
rate suppression derive from lateral (spatial or magnetic) confinements which are strong enough
to disregard PZ interaction.

\subsection{Vertically Coupled Quantum Dots}

\begin{figure}[p]
\includegraphics[width=8.5cm,clip]{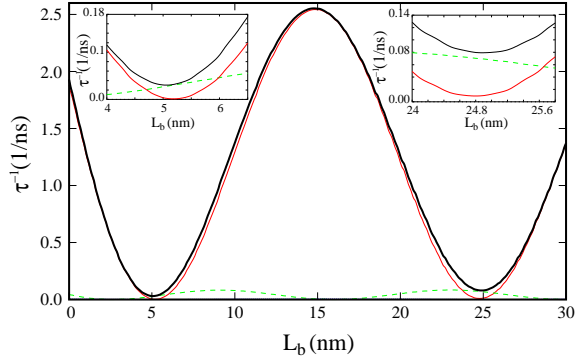}
\caption{(Color online) Acoustic phonon scattering rate as a function of the
barrier width $L_b$ for the electron transition $(n,m,g)=(0,1,0)
\rightarrow (0,0,0)$ in a CQD structure with $\hbar \omega_0=1$ meV
and $L_z=5$ nm, . The insets zoom in the regions around the DP 
scattering minima. Scattering mechanisms are represented as in the
legend of Figure \ref{Fig1}.}\label{Fig4}
\end{figure}

CQDs constitute an interesting system for phonon scattering
modulation because they allow the electron charge to distribute
among the various quantum wells along the vertical direction. As a
consequence, the total height of the artificial molecule, as CQDs are
often called, replaces the SQD height as the critical parameter
leading to relaxation rate suppressions when it is in-phase
with the phonon wavelength. It is then possible to minimize the
scattering rate of the $(0,1,0) \rightarrow (0,0,0)$ (intradot)
transition by proper structure design\cite{ZanardiPRL} or using
magnetic fields\cite{BertoniAPL,BertoniPE} even when the lateral
confinement is so weak that the suppression described above for SQDs
cannot occur. In particular, it has been predicted that for periodic
values of the interdot barrier thickness, $L_b$, decoherence-free systems may
be built.\cite{ZanardiPRL} This prediction was obtained considering
only DP scattering, and one may wonder whether this picture holds
when we include PZ interactions. To this end, in Figure \ref{Fig4} we
plot the scattering rate of two CQDs with $L_z=5$ nm and $\hbar
\omega_0=1$ meV, as a function of the barrier thickness. The total
scattering rate oscillates periodically, with most of the
contribution coming from DP interactions. Nevertheless, the TA-PZ
contribution becomes predominant at the values of $L_b$ where
complete suppression of LA-DP scattering is predicted (see insets around
$L_b=5$ nm and $L_b=25$ nm in Figure 4); indeed, the DP and PZ mechanism have minima
at different $L_b$ due to the difference in the sound velocity and,
therefore, in the wavelength of the emitted phonons. This imposes a
limit on the barrier-thickness-induced suppression of intradot
transition rates reported in Ref.~\onlinecite{ZanardiPRL}, which can
be nonetheless minimized by growing CQDs with stronger lateral
confinement, so that PZ interaction strength rapidly decreses (see
Figure \ref{Fig1}).

\begin{figure}[p]
\caption{(Color online) Acoustic phonon scattering rate as a function of the
magnetic field for the electron transition $(n,m,g)=(0,1,0)
\rightarrow (0,0,0)$ in two CQD structures with $L_z=12$ nm, $L_b=5$
nm and different lateral confinement. The arrows point to the position
of the relaxation rate minima expected including the DP interaction
only (Refs. \onlinecite{BertoniAPL,BertoniPE})}\label{Fig5}
\includegraphics[width=8.5cm,clip]{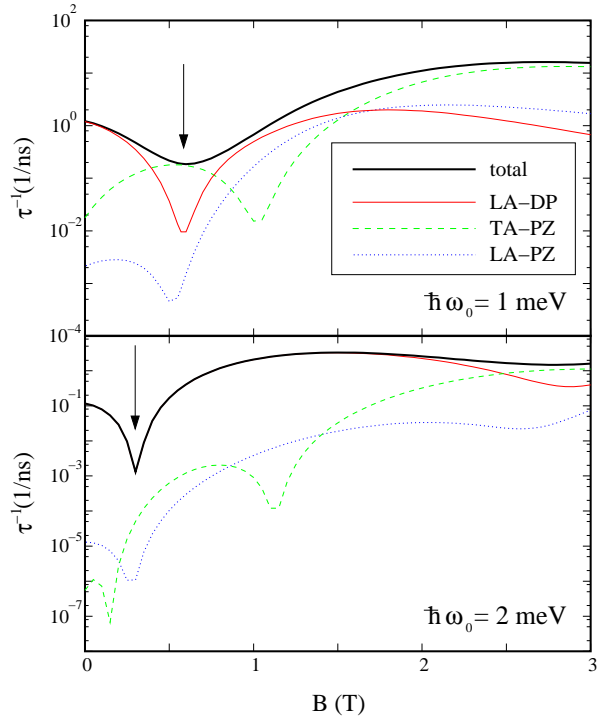}
\end{figure}

Recently we have shown that $B$-induced control of the DP scattering rate in CQDs is also
possible.\cite{BertoniAPL,BertoniPE}
Again, the question arises concerning the influence of PZ scattering
in this picture. Figure \ref{Fig5} illustrates the relaxation rate for
two CQD structures with $L_z=12$ nm, $L_b=5$ nm and different
lateral confinement. The arrows point to the position of the
expected minima according to the DP
description.\cite{BertoniAPL,BertoniPE} One can see that for $\hbar
\omega_0=1$ meV the minimum of DP scattering coincides with the
maximum of TA-PZ scattering, and therefore the expected suppression
of the relaxation rate is greatly inhibited. Yet, this no longer
occurs for $\hbar \omega_0=2$ meV, where PZ interactions are less
relevant at weak fields due to the stronger spatial lateral
confinement.
 It is worth stressing that this suppression is found for a lateral confinement strength which is
too weak for the SQD-based suppression to be efficient [see $\hbar
\omega_0=2$ meV curve in Figure \ref{Fig2}(a)]. On the other hand,
simultaneous control of DP and TA-PZ scattering rates, aimed at
making their respective minima coincide, is an intrincate process
which follows from the interplay among QD geometry, composition and
growth direction. Therefore, we conclude that CQDs stand as a firm
alternative to SQDs for control of intradot transitions phonon
relaxation only when both scattering sources have significantly
different weights. This can be achieved for instance using QDs with
lateral confinement $\sim 2$ meV, or building QDs with narrow-gap
materials where the spacing between conduction band states is
increased and hence
PZ influence is smaller.\\

\begin{figure}[p]
\includegraphics[width=8.5cm,clip]{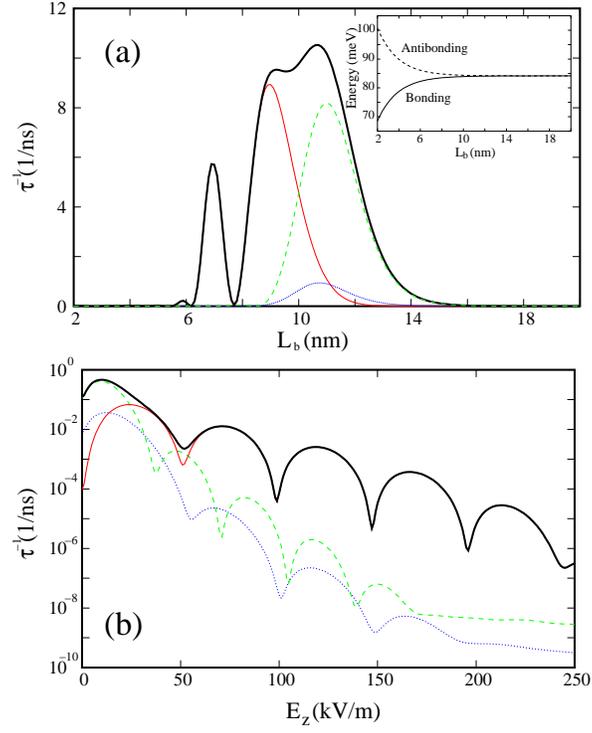}
\caption{(Color online) Acoustic phonon scattering rate as a function of (a) the
barrier width $L_b$, and (b) the electric field, for the electron
transition $(n,m,g)=(0,0,1) \rightarrow (0,0,0)$ in a CQD structure
with $\hbar \omega_0=5$ meV and $L_z=5$ nm. Inset of panel (a):
energy of the lowest bonding (solid line) and antibonding (dashed
line) states vs barrier width. In panel (b) $L_b=15$ nm.
Scattering mechanisms are represented as in the legend of Figure
\ref{Fig1}.}\label{Fig6}
\end{figure}

Next, we analyze the transition rate between the lowest bonding and
antibonding states of two CQDs, $(n,m,g)=(0,0,1) \rightarrow
(0,0,0)$ (interdot or isospin transition). The interest of this
transition is in part connected with the demand of coherent
tunneling between the two dots for isospin-based implementations of
quantum gates\cite{BayerSCI}, as well as with quantum dot cascade
laser devices performance.\cite{WingreenIEEE} The wavelength of the
emitted phonon is in this case controlled by the tunneling energy,
which can be tuned by means of either the barrier thickness or an
electric field applied along the $z$ direction
($E_z$).\cite{BertoniAPL} In Figure \ref{Fig6} we study the effect of
both mechanisms for CQD devices with $L_z=5$ nm and $\hbar
\omega_0=5$ meV. Figure \ref{Fig6}(a) represents the transition rate
vs the barrier width. The total rate depends strongly on $L_b$.
However, only a few non-periodic oscillations are
observed, as opposed to the intradot transition case (Figure
\ref{Fig4}). This is due to the changing tunneling energy (see
inset), which makes the phonon wavelength vary exponentially with
$L_b$. On the other hand, we note that PZ interactions become
dominant when the barrier thickness is large, owing to the small
tunneling energy. These results can be related to recent experiments
which reported different relative photoluminescence intensities for
bonding and antibonding exciton states in a pair of CQDs as a
function of the interdot distance.\cite{OrtnerPRB} Since the
intensity of photoluminescence peaks is proportional to the lifetime
of conduction states, and acoustic phonon coupling was identified as
the main relaxation source in the experiment, these observations
indicated a non-monotonic dependence of the antibonding state
lifetime on the barrier thickness, in agreement with
Figure \ref{Fig6}(a). In Figure \ref{Fig6}(b) we depict the
$E_z$-dependent scattering rate corresponding to CQDs with $L_b=15$
nm. As in previous works\cite{BertoniAPL}, we find
order-of-magnitude oscillations of the DP scattering rate, which
also take place for PZ rates albeit with different period for the TA
phonons. Interestingly, the PZ scattering rate decay is much faster
than that of the DP, due to the different wave vector dependence of
the matrix element. As a result, even if PZ prevails in the absence
of external fields, the use of an electric field soon turns DP into
the dominant relaxation mechanism, and one can think of
electric-field-induced suppressions of the scattering rate in terms
of DP interaction only. We would like to point out that the interdot
transition rate minima observed in Figure \ref{Fig6} have no
counterpart in laterally coupled dots.\cite{StavrouPRB} This
represents a fair advantage of vertically coupled structures with
regard to quantum gate implementation schemes, since it allows one
to reduce charge decoherence times dramatically. The underlying
reason for this difference lies in the larger total vertical
dimension of the vertically coupled double QD system, which makes it
possible to reach in-phase relation between the electron
wavefunction and the phonon plane wave along $z$ for small tunneling
energies (i.e., long phonon wavelengths).

\begin{figure}[p]
\includegraphics[width=8.5cm,clip]{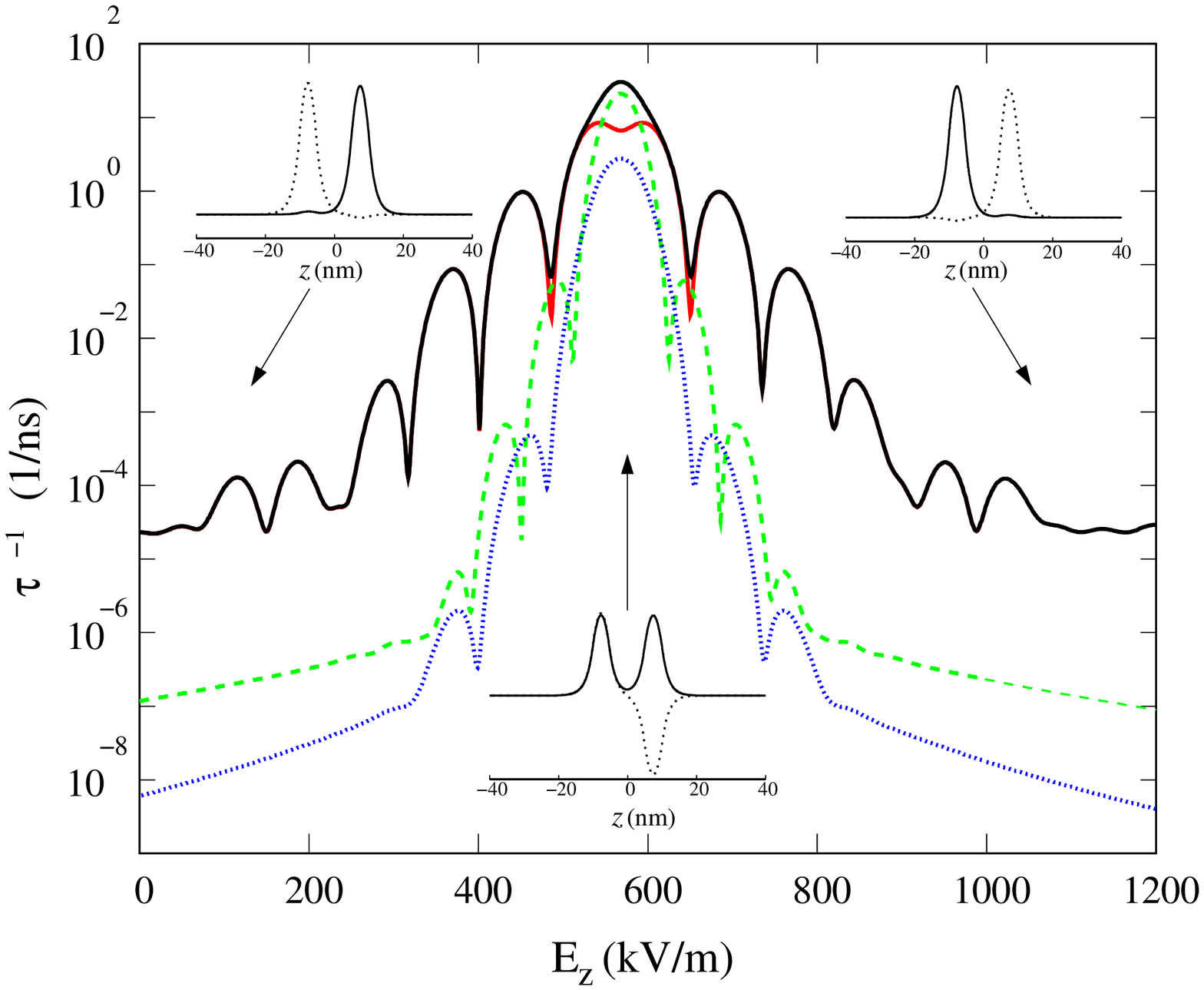}
\caption{(Color online) Acoustic phonon scattering rate as a function of the
electric field for the electron transition $(n,m,g)=(0,0,1)
\rightarrow (0,0,0)$ in a heteronuclear CQD structure, with $L_b=10$
nm, $\hbar \omega_0=5$ mev, $L_z=5$ nm (bottom QD) and $L_z=5.5$ nm
(upper QD). Insets: electron wavefunction (arbitrary units) in the
double quantum well for $E_z=0$ kV/cm (left), $E_z=568$ kV/m
(center) and $E_z=1200$ kV/m (right), with solid (dotted) lines
representing the lowest bonding (antibonding) state. Scattering
mechanisms are represented as in the legend of Figure
\ref{Fig1}.}\label{Fig7}
\end{figure}

So far we have considered CQDs composed by two identical dots (in
analogy with atomic molecules, we refer to these systems as
'homonuclear'). However, in practice this is hard to achieve and the
size of the coupled dots is usually slightly different, leading to
'heteronuclear' systems.\cite{PiPRL,LedentsovPRB} Molecular levels
are very sensitive to small changes in the size of the coupled
dots,\cite{PiPRL,FonsecaPRB} so that one may also expect significant
effects on the transition rates. To analyze this case, in
Figure \ref{Fig7} we consider a CQD structure with $L_b=10$ nm. Both
QDs have the same lateral confinement $\hbar \omega_0=5$ meV, but
the bottom well width is $L_z=5$ nm, while the upper well one is $L_z=5.5$ nm.
The effect of an electric field on this heteronuclear system is to
introduce order-of-magnitude oscillations which first lead to an
enhancement of the scattering rate and then to a reduction in a
clearly symmetric fashion. These results are interpreted as follows.
In the absence of external fields, bonding and antibonding states
(solid and dotted lines in the insets, respectively) localize in
opposite wells. As a consequence, the factor $\langle \Psi_f|\,e^{-i
\mathbf{q r}}\,| \Psi_i \rangle$ in Eq.~(\ref{eq1}) is reduced as
compared to the homonuclear case, where the charge density of
bonding and antibonding states are equally distributed between the
dots. For a given electric field we can compensate for the different
well width of the two QDs, and force an identical charge density
distribution of bonding and antibonding states.  This maximizes
$\langle \Psi_f|\,e^{-i \mathbf{q r}}\,| \Psi_i \rangle$ and
therefore the relaxation rate. Finally, for stronger fields the
original localization of states is reversed. It is also worth noting
that the total scattering rate in Figure \ref{Fig7} is essentially
given by the DP interaction, except in the vicinity of the
'homonuclearity' point, where the energy of bonding and antibonding
levels are very close and TA-PZ relaxation prevails. For practical
purposes, one is often interested in obtaining coherent
delocalization of the particles between the two dots, i.e. reaching
a homonuclear system with long electron lifetimes.\cite{BayerSCI}
Figure \ref{Fig7} shows that a relatively small departure from the
electric field which gives the homonuclear charge distribution, from
$E_z=570$ to $E_z=485$ kV/m, increases the electron lifetime by
three orders of magnitude, while partially preserving electron
delocalization.

\section{Summary}

Single-electron relaxation rates in GaAs SQDs and CQDs due to
coupling with acoustic phonons have been investigated. PZ
interactions due to TA phonons, often disregarded in the literature, 
constitute the major source of scattering for small ($< 0.5$ meV)
electron transition energies. For larger gaps, DP interactions due
to LA phonons prevail. Indeed, we identify many situations where PZ effects
become critical, such as SQDs subjected to an axial magnetic field
or the isospin transition in CQDs with large interdot distances.
Nevertheless, we have shown that proper structure design makes it
possible to control the scattering rates of both SQDs and CQDs by
orders of magnitude. Suppression of the scattering rate in SQDs can
be attained for moderately large (few meV) lateral confinement,
while CQD structures permit to extend it down to fairly weak lateral
confinements. Similar results can be achieved using external fields
even when the structures do not have the optimum geometry. Unlike
laterally CQDs, vertically CQDs can be used to obtain enhanced
electron lifetimes (in the range of microseconds) in excited
'molecular' states, which has important implications for quantum
computation devices. These results are found to be robust for
size-mismatched CQDs.

\begin{acknowledgments}
Financial support from the Italian Ministry of Foreign Affairs (DGPCC)
 the Italian Ministry of Research under the MIUR-FIRB (RBIN04EY74) 
program and CNR-INFM Calcolo Parallelo 2005 and 2006 is acknowledged.
This work has been supported in part by the EU under the TMR network 
``Exciting'' (J.I.C.).
\end{acknowledgments}

\clearpage

\pagebreak

\pagebreak

\pagebreak

\pagebreak

\pagebreak

\pagebreak

\begin{thebibliography}{26}
\bibitem{FujisawaNAT}
T. Fujisawa, D.G. Austing, Y. Tokura, Y. Hirayama, and S. Tarucha, Nature (London) {\bf 419}, 278 (2002).

\bibitem{FujisawaJPCM}
T. Fujisawa, D.G. Austing, Y. Tokura, Y. Hirayama, and S. Tarucha, J. Phys.: Cond. Matter {\bf 15}, R1395 (2003).

\bibitem{FujisawaSCI}
T. Fujisawa, T.H. Oosterkamp, W.G. van der Wiel, B.W. Broer, R. Aguado, S. Tarucha, and L.P. Kouwenhoven,
Science {\bf 282}, 932 (1998).

\bibitem{OrtnerPRB}
G. Ortner, R. Oulton, H. Kurtze, M. Schwab, D.R. Yakovlev, M. Bayer, S. Fafard, Z. Wasilewski, and
P. Hawrylak, Phys. Rev. B {\bf 72}, 165353 (2005).

\bibitem{ZanardiPRL}
P. Zanardi, and F. Rossi, Phys. Rev. Lett. {\bf 81}, 4752 (1998);
P. Zanardi, and F. Rossi, Phys. Rev. B {\bf 59}, 8170 (1999).

\bibitem{SugawaraAPL}
M. Sugawara, K. Mukai, and H. Shoji, Appl. Phys. Lett. {\bf 71}, 2791 (1997).

\bibitem{WingreenIEEE}
N.S. Wingreen, and C.A. Stafford, IEEE J. Quantum Electron. {\bf 33}, 1170 (1997).

\bibitem{BockelmannPRB}
U. Bockelmann, Phys. Rev. B {\bf 50}, 17271 (1994).

\bibitem{BertoniAPL}
A. Bertoni, M. Rontani, G. Goldoni, F. Troiani, and E. Molinari, Appl. Phys. Lett. {\bf 85}, 4729 (2004).

\bibitem{BertoniPE}
A. Bertoni, M. Rontani, G. Goldoni, F. Troiani, and E. Molinari, Physica E (Amsterdam) {\bf 26}, 427 (2005).

\bibitem{BertoniPRL}
A. Bertoni, M. Rontani, G. Goldoni, and E. Molinari, Phys. Rev.
Lett. {\bf 95}, 066806 (2005).

\bibitem{ChengPRB}
J.L. Cheng, M.W. Wu, and C. L\"u, Phys. Rev. B {\bf 69}, 115318 (2004).

\bibitem{WuPRB}
Z.J. Wu, K.D. Zhu, X.Z. Yuan, Y.W. Jiang, and H. Zheng, Phys. Rev. B {\bf 71}, 205323 (2005).

\bibitem{StavrouPRB}
V.N. Stavrou, and X. Hu, Phys. Rev. B {\bf 72}, 075362 (2005).

\bibitem{weakdots}
S. Tarucha, D.G. Austing, S. Sasaki, Y. Tokura, W. van der Wiel, L.P. Kouwenhoven, Appl. Phys. A {\bf 71}, 367 (2000); U. Bockelmann, Ph. Roussignol, A. Filoramo, W. Heller, G. Abstreiter, K. Brunner, G. B\"ohm, and G. Weimann, Phys. Rev. Lett. {\bf 76}, 3622 (1996); J. Kyriakidis, M. Pioro-Ladriere, M. Ciorga, A.S. Sachrajda, and P. Hawrylak, to be published in Phys. Rev. Lett. (cond-mat/0111543).

\bibitem{ReimannRMP}
S. M. Reimann and M. Manninen, Rev. Mod. Phys. 74, 1283 (2002).

\bibitem{approx}
The use of bulk phonons is a reasonable approximation for GaAs/Al$_{1-x}$Ga$_x$As heterostructures,
owing to the similar lattice constants of the low-dimensional structure and embedding matrix materials.
However, if the materials were elastically dissimilar, confined acoustic phonon modes should be
considered, see e.g. A.A. Balandin, J. Nanosci. Nanotech. {\bf 5}, 1015 (2005), and references therein.

\bibitem{Gantmakher_book}
V.F. Gantmakher, and Y.B. Levinson, \emph{Carrier Scattering in Metals and Semiconductors}, (Modern Problems in Condensed Matter Sciences, vol. 19, Elsevier Science Publishers, 1987).

\bibitem{ZookPR}
J.D. Zook, Phys. Rev. {\bf 136}, A869 (1964).

\bibitem{Tin_book}
C.S. Ting (ed.), \emph{Physics of Hot Electron Transport in Semiconductors}, (World Scientific, 1992).

\bibitem{Landolt_book}
O. Madelung (Ed.), \emph{Landolt-B\"ornstein, Numerical Data and Functional Relationships
in Science and Technology, Vol. 17 Semiconductors, Group IV Elements and III-V Compounds}, (Springer-Verlag, 1982).

\bibitem{FlorescuPE}
M. Florescu, S. Dickman, M. Ciorga, A. Sachrajda, and P. Hawrylak, Physica E (Amsterdam) {\bf 22}, 414 (2004);
M. Florescu, and P. Hawrylak, Phys. Rev. B {\bf 73}, 045304 (2006). 

\bibitem{BayerSCI}
M. Bayer, P. Hawrylak, K. Hinzer, S. Fafard, M. Korkusinski, Z.R. Wasilewski, O. Stern, and A. Forchel, Science {\bf 291}, 451 (2001).

\bibitem{PiPRL}
M. Pi, A. Emperador, M. Barranco, F. Garcias, K. Muraki, S. Tarucha, and D. G. Austing, Phys. Rev. Lett. {\bf 87}, 066801 (2001).

\bibitem{LedentsovPRB}
N.N. Ledentsov, V.A. Shchukin, M. Grundmann, N. Kirstaedter, J. B\"ohrer, O. Schmidt, D. Bimberg, V.M. Ustinov,
A. Yu. Egorov, A.E. Zhukov, P.S. Kop'ev, S.V. Zaitsev, N.Yu. Gordeev, Zh.I. Alferov,
A.I. Borovkov, A.O. Kosogov, S.S. Ruvimov, P. Werner, U. G\"osele, and J. Heydenreich, Phys. Rev. B {\bf 54}, 8743 (1996).

\bibitem{FonsecaPRB}
L.R.C. Fonseca, J.L. Jimenez, and J.P. Leburton, Phys. Rev. B {\bf 58}, 9955 (1998).

\end{thebibliography}
\end{document}